\pdfoutput=1
\documentclass{article}

\PassOptionsToPackage{numbers, compress}{natbib}
\usepackage[final,dblblindworkshop]{neurips_2026}
\workshoptitle{Agents in the Wild}

\usepackage[utf8]{inputenc}
\usepackage[T1]{fontenc}
\usepackage{hyperref}
\usepackage{url}
\usepackage{booktabs}
\usepackage{amsfonts}
\usepackage{amsmath}
\usepackage{amssymb}
\usepackage{nicefrac}
\usepackage{microtype}
\usepackage{xcolor}
\usepackage{graphicx}
\usepackage{enumitem}
\graphicspath{{./}{../../scenarios/}}
\usepackage{xspace}
\newcommand{\tool}{\textsc{AgentCap}\xspace}

\title{\tool: Capability-Based Runtime Authorization for Secure LLM Agent Execution}

\author{%
  Aravinda Kashyap Kabbinamakki Suresh \\
  Virginia Tech \\
  Blacksburg, VA, USA \\
  \texttt{aravindakashyap@vt.edu} \\
  \And
  Peng Gao \\
  Virginia Tech \\
  Blacksburg, VA, USA \\
  \texttt{penggao@vt.edu} \\
}

\begin{document}
\maketitle
\begin{abstract}

Language-model agents commonly collapse two decisions that security systems
separate: selecting an action and determining whether that action is
authorized. When a model exercises standing authority, indirect prompt
injection can convert influence over its reasoning into unauthorized effects.
We present \tool, a runtime authorization layer that externalizes this
decision from the model and mediates every tool invocation through an MCP
wrapper. Each request declares an intent, and \tool evaluates policy
constraints over trust, workflow prerequisites, mutually exclusive authority,
and the provenance of security-sensitive arguments. Authorized requests receive
short-lived, budget-bounded capability tokens that are validated at the tool
boundary. Delegation is restricted to session-bound parent--child chains that
preserve action and resource while budget and lifetime can only be attenuated, preventing
authority from being expanded or reused across unrelated agent sessions. A
declarative policy language provides a common representation for runtime
enforcement and predeployment analysis. Across AgentDojo and InjecAgent, \tool reduces
attack success in every evaluated model--benchmark setting, including zero
attack success with GPT-5.4 on both benchmarks, while utility under attack
remains within 4.18 percentage points of the native agents. Controlled
adversarial scenarios further show that \tool blocks every attempted
unauthorized execution (ASR 0\%) while preserving all legitimate operations exercised.
These results demonstrate that agent security requires authorization of
model-proposed effects at the tool boundary, independently of the model's
reasoning.
\end{abstract}
\section{Introduction}
\label{sec:intro}

Tool-using language-model agents collapse a distinction that conventional security architectures treat as fundamental: the separation between proposing an action and authorizing its execution. An agent selects tool calls while operating with credentials that often remain valid throughout an entire session. Its decisions are simultaneously conditioned on potentially adversarial content, including retrieved documents, messages, web pages, and prior tool outputs. Consequently, an indirect prompt injection need not bypass an explicit authorization check. It can instead redirect an already authorized agent toward an attacker-selected action \citep{greshake2023signedup,debenedetti2024agentdojo,asb2025}. This coupling between model inference and ambient authority turns failures of instruction following into security violations.

Existing defenses address important parts of this problem but do not by themselves establish a per-invocation authorization boundary. Prompt filtering and model alignment attempt to prevent untrusted content from influencing behavior, yet their decisions remain probabilistic and operate within the reasoning process targeted by the attacker. Identity- and role-based policies restrict the tools available to an agent, but commonly grant authority at session scope rather than for a particular task or invocation. Expiration and rate limits restrict when and how often credentials may be used, but cannot distinguish a legitimate call from an attacker-directed call made within the same limits. Recent systems therefore move enforcement outside the model through deterministic call policies, task-specific permission structures, and information-flow controls \citep{progent2025,miniscope2025,fides2025,clawguard2026}. These developments motivate a stronger principle: every model-generated action should be treated as an untrusted request whose authority must be established independently.

We present \tool, a capability-based authorization layer for AI agents. Capability systems attach authority to an unforgeable object that denotes a specific permission rather than to the ambient identity of its holder, an abstraction introduced to address confused-deputy failures \citep{hardy1988deputy}. \tool mediates every proposed tool invocation through a deterministic authorization gate placed between the agent and its tools. The gate evaluates the request independently of the model and forwards it to the underlying tool only when the applicable authorization conditions are satisfied. An approved request receives a short-lived signed capability scoped to an action, resource, session, declared intent, and invocation budget. A separate gateway verifies this capability before forwarding the call to the underlying tool. Action selection consequently remains with the model, while authority is determined and enforced outside it.

Capability scoping alone is insufficient when an attacker manipulates the arguments of an otherwise legitimate operation. \tool therefore incorporates parameter-level provenance into authorization. The policy associates content-producing tools and selected output fields with integrity labels, then checks whether security-sensitive arguments are supported by trusted task context. Consider an injected transaction record that instructs the agent to transfer funds to an attacker. Although the resulting payment may carry a valid high-level intent, its recipient originates only from untrusted content. The provenance check rejects this instantiation of the operation without prohibiting legitimate payments authorized by the user. This design extends recent information-flow approaches that distinguish trusted instructions from lower-integrity data \citep{fides2025,rtbas2025}.

The authorization state also evolves during execution. Suspicious requests and
execution outcomes update a per-agent trust state, with threshold crossings
triggering capability revocation or sandboxing. Delegation across agents is
represented as a signed attenuation chain: each hop preserves the authorized
action and resource, never increases the lifetime or budget, and strictly
reduces the remaining delegation depth. Capabilities remain bound to their agent lineage and session, preventing agents in unrelated sessions from reusing or extending delegated
authority. A compact declarative policy language specifies capability grants, prerequisites, mutually exclusive permissions, delegation constraints, trust responses, provenance requirements, and valid workflows. Because these relationships are explicit, a static analyzer can detect policy interactions before deployment.

\paragraph{Contributions.}
This work makes three major contributions: 
\begin{itemize}[leftmargin=*]
    \item \textbf{Capability-based authorization for single- and multi-agent
    execution.}
    We introduce signed, agent- and session-bound capabilities that constrain
    tool authority by action, resource, intent, lifetime, and budget. Authority
    may cross agents only through explicit parent-to-child delegation, with
    attenuation preventing a child from exceeding the authority of its parent
    or reusing authority from an unrelated session.

    \item \textbf{An MCP-integrated enforcement architecture.}
    We implement \tool as an MCP wrapper and authorization gateway. The
    wrapper obtains request-scoped authority for each proposed invocation, and
    the gateway forwards the call only after validating the resulting
    capability. A declarative policy language unifies capability issuance,
    workflow prerequisites, mutually exclusive permissions, argument
    provenance, trust-triggered revocation, and delegation constraints.

    \item \textbf{Static and empirical validation.}
    A surface analyzer detects over- and under-permissive policy interactions
    before deployment. Across the complete AgentDojo suite and
    InjecAgent, \tool reduces attack success in every evaluated
    model--benchmark setting and reaches zero attack success with GPT-5.4 on
    both benchmarks. Controlled single- and multi-agent scenarios additionally
    block every attempted unauthorized execution while preserving all
    legitimate operations exercised.
\end{itemize}

\section{Threat Model}
\label{sec:threat}

We consider single- and multi-agent systems in which an adversary may control retrieved content, tool outputs, or inter-agent messages and thereby cause the model to propose attacker-selected tool calls and arguments. The model is not trusted to interpret untrusted content correctly or enforce authorization policy. We trust \tool's policy engine, MCP wrapper, authorization gateway, cryptographic keys, and underlying transport; the adversary cannot forge signatures or bypass the mediated tool interface. Our security objective is to ensure that a compromised agent can exercise only the authority encoded in a valid capability, and that authority crosses agent boundaries only through policy-authorized, attenuating parent-to-child delegation.

\tool treats the policy, provenance labels, and outcome labels as
security-critical configuration supplied by the deployment environment, not as
facts inferred from the model. This makes the trust boundary explicit and
reviewable: a policy author must correctly identify the actions a task may
perform, the sources that may justify protected arguments, and the outcomes
that warrant containment. The policy compiler, surface analyzer, and audit log
support review of these artifacts, but cannot compensate for an incorrectly
trusted source or an over-permissive policy. Under these assumptions,
\tool's guarantee is deterministic: a proposed effect reaches a tool only
when it satisfies the configured authorization policy.

\section{The \texorpdfstring{\tool}{AgentCap} Architecture}
\label{sec:architecture}

The \tool architecture is organized around a single security invariant: no model-generated tool call reaches an MCP server without independently authorized and verifiable authority. \tool separates action proposal, capability issuance, and capability verification into distinct trust domains. The MCP wrapper serves as the agent's tool interface, while the policy engine determines whether a requested action is admissible and issues a narrowly scoped capability when authorization succeeds. Capabilities are bound to the agent and session for which they are issued, thereby isolating the authority available to concurrent agent sessions. Within a multi-agent session, a parent may delegate only an attenuated subset of its authority to a child, and agents outside that delegation lineage cannot exercise it, even if the orchestration layer permits them to exchange messages. Communication therefore does not imply authority: an agent may provide information to another agent, but any subsequent access to or modification of external resources remains subject to the receiving agent's own capabilities. The authorization gateway verifies these constraints immediately before forwarding an invocation to the underlying MCP server.

\begin{figure}[t]
  \centering
  \includegraphics[width=\linewidth]{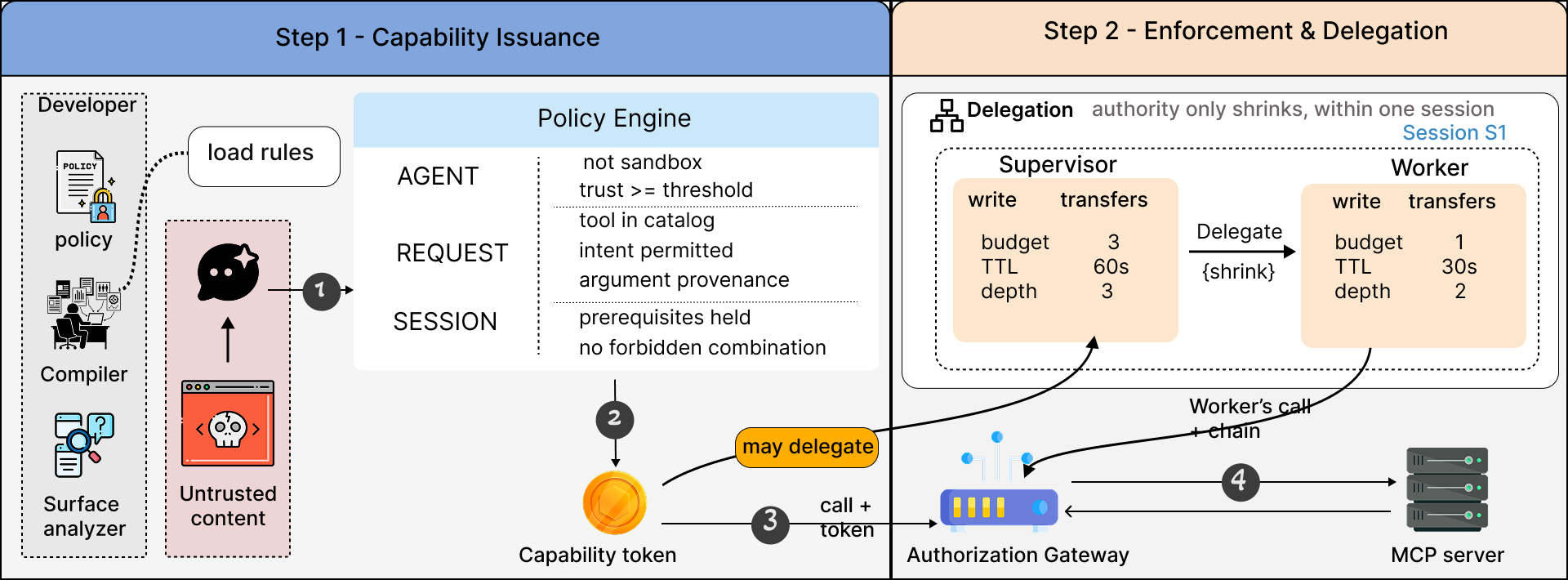}
    \caption{The \tool architecture. \emph{Step~1 (capability issuance):} the
    developer's policy is compiled into the rules loaded by the policy engine,
    and the surface analyzer checks the same policy before deployment. The
    agent, which may be steered by untrusted content, proposes a tool call with a
    declared intent~(1). The policy engine checks the agent (not sandboxed,
    trust at or above its threshold), the request (tool in the catalog, intent
    permitted, argument provenance), and the session (prerequisites held, no
    forbidden combination). Only if every check passes does it issue a signed,
    short-lived capability token~(2); otherwise the call is denied.
    \emph{Step~2 (enforcement and delegation):} the agent presents the token
    with its call~(3), and the authorization gateway verifies the token before
    forwarding the call to the MCP server and returning the result~(4). A token
    holder may delegate it within the same session: the worker's token keeps the
    supervisor's action and resource, while its budget, lifetime (TTL), and
    remaining delegation depth can only shrink. The gateway accepts the worker's
    call only after verifying its delegation chain back to the policy-issued
    token.}
  \label{fig:architecture}
\end{figure}

\subsection{Capability Tokens}
\label{sec:tokens}
A capability token is a tuple
\[
  \tau =
  \langle\, j,\; u,\; a,\; r,\; i,\; s,\; \theta,\; q,\; e,\; b,\; P,\; G,\; \rho \,\rangle,
\]
binding a token identifier $j$, agent $u$, action $a$, resource $r$, declared
intent $i$, session $s$, trust score $\theta$ at issuance, issuance time $q$,
expiry $e$, initial call budget $b$, prerequisite set $P$, deny-combination
groups $G$, and audit-record reference $\rho$. The policy engine signs these
claims using Ed25519 and encodes the resulting token as a JWT. A verifier can
authenticate the signed claims without contacting the issuer. Before execution,
the authorization gateway additionally checks expiry, revocation, remaining budget,
presenting-agent identity, and agreement with the requested action and
resource. Thus, the token protects immutable authorization claims, while the
gateway combines those claims with current enforcement state.

A delegation token additionally identifies its delegator, delegatee, parent
token, remaining redelegation depth, and optional behavioral bound. For a child
authorization $\tau'$ derived from a parent authorization $\tau$, we write
$\tau' \sqsubseteq \tau$ when
\[
  a' = a, \qquad r' = r, \qquad b' \leq b, \qquad
  e' \leq e, \qquad d' < d,
\]
where $d'$ and $d$ denote the remaining redelegation depths associated with the
child and parent, respectively. The relation is transitive, and the strict
decrease in depth bounds the length of every delegation chain. A delegated
token may be derived from its parent only when these constraints hold.

\subsection{Policy-Governed Capability Issuance}
\label{sec:policyengine}

The policy engine maps a proposed tool call to a capability decision. Each
request identifies an agent, session, capability, declared intent, and set of tool parameters. Issuance requires seven conditions, evaluated in order. The agent must not be sandboxed, the requested tool must resolve to a capability in the catalog, protected parameters must have acceptable provenance, and the declared intent must appear in that capability's intent whitelist. The agent must then meet the capability's trust threshold and already hold, within the same session, a valid token for every prerequisite capability. Finally, issuance must not complete a forbidden combination (\S\ref{sec:policylang}) with capabilities already held in that session. The checks are evaluated sequentially to fail closed, but their semantics are conjunctive: a token is
issued only when every condition holds, and it binds the resulting authority to the approved action, resource, intent, agent, session, lifetime, and call budget. The declared intent is supplied by the model and is therefore not a trusted assertion of benign behavior. Rather, it binds a call to one of the purposes for which the policy has pre-authorized the capability, so a token issued for one declared purpose cannot be reused under an unrelated one. A denial at any stage produces an audit record and decays the agent's trust score, so an agent probing for a gap in the policy is observable while it probes.

\paragraph{Argument provenance.}
Policy authors designate the task instruction and selected application-owned fields as trusted context, and designate free-form retrieved or tool-produced fields as untrusted. A \texttt{bound\_params} declaration attaches this provenance check to a security-sensitive argument - for example, a payment recipient or a credential. During execution, \tool records the values made available through each source and rejects a bound argument when its support derives solely from untrusted content. The check is therefore neither a classifier of the model's intent nor a blanket ban on values that appear in tool output: a value independently present in trusted task context remains admissible. This separation lets the policy permit a user-requested payment while rejecting an attacker-supplied recipient embedded in a document the agent read.

\subsection{A Declarative Policy Language}
\label{sec:policylang}

The authorization procedure is parameterized by a declarative policy comprising five block types. A \texttt{grant} defines a capability's action, resource, admissible intents, trust threshold, prerequisites, budget, lifetime, and exclusion groups. A \texttt{deny\_combination} prohibits specified groups from being held together. A \texttt{delegation} identifies permitted delegator and delegatee types and bounds the resulting chain. An \texttt{on\_trust} block maps trust thresholds to revocation actions, while a \texttt{workflow} describes the capabilities required by a legitimate task. The first four blocks govern runtime enforcement; workflow declarations are consumed only by the static analyzer.

\begin{figure}[t]
\centering
\scriptsize
\begin{verbatim}
grant send_money {
  action: write
  resource: "bank_account/transfers"
  intent: [pay_bill, send_payment]
  trust: >= 0.8
  requires: [get_balance]
  ttl: 30s
  budget: 1
  deny_groups: [financial_write]
  bound_params: [recipient]
}

delegation supervisor_to_worker {
  from: type(supervisor)
  to: type(worker)
  max_depth: 3
  require: attenuate_only
}

on_trust agent below 0.3 {
  revoke: all
  sandbox: true
}
\end{verbatim}
\caption{Representative blocks from the \tool banking policy. A capability grant constrains issuance, a delegation rule restricts authority transfer between agent types, and a trust rule revokes authority when trust falls below its configured threshold. Appendix~\ref{app:grammar} gives the complete grammar.}
\label{fig:policylang}
\end{figure}

The compiler lowers these blocks directly into the runtime objects used by the policy engine, delegation manager, trust manager, and surface analyzer. There is no separate policy evaluator or intermediate enforcement path. This common representation allows the analyzer to reason jointly about prerequisites, exclusions, and legitimate workflows. As reported in \S\ref{sec:analysisresults}, it identifies capabilities rendered unreachable by inconsistent constraints, while regression tests confirm that compiled grants and exclusions reproduce their hand-written configurations.

\subsection{Delegation as Attenuated Chains}
An agent holding a capability token may delegate a bounded portion of its authority to another agent. Delegation admits a child token that is an attenuation of the parent, in the sense of Section~\ref{sec:tokens}. A request that would widen authority is malformed, not merely denied. Each hop is signed by the delegator over every field except the signature itself.

Verification walks the chain from leaf to root. It checks each hop's signature, its attenuation against a snapshot of its parent's constraints taken at delegation time, and revocation at every hop, before verifying the root token's signature against the policy engine's key. A chain is valid only if every hop and the root are simultaneously valid. Verification never trusts a cached ``was valid'' state.

A \texttt{delegation} rule governs which agent types may delegate to which. Each hop requires an explicit rule for its delegator and delegatee types. A requested depth exceeding that rule's maximum is rejected. The rule's default behavioral bound applies when the delegator supplies none.

\subsection{Trust-Driven Containment}
\label{sec:trust}

\tool maintains a trust score $t \in [0,1]$ for each agent. A normal outcome updates the score as $t' = t + r(1-t)$, moving it toward $1$, while an anomalous outcome applies the multiplicative decay $t' = t(1-d)$. Policy-defined thresholds progressively restrict the agent's authority: crossing the first threshold revokes sensitive capabilities, while crossing the second revokes all capabilities and sandboxes the agent. Anomalous outcomes include both denials by the gate (\S\ref{sec:policyengine}) and executed calls that the host environment labels anomalous.

We use $t_0=0.8$, $d=0.3$, and $r=0.01$ as the default configuration. Under the decay rule, three consecutive anomalies produce trust scores of $0.560$, $0.392$, and $0.274$. The agent consequently loses write authority after the second anomaly and is sandboxed after the third. We evaluate these choices through a controlled one-factor-at-a-time sweep over the initial trust, decay rate, recovery rate, and both revocation thresholds. Across the 23 evaluated configurations, sandboxing occurred after 1 to 10 anomalous outcomes, and the legitimate capability request included in each trial remained authorized. Full results are reported in Appendix~\ref{app:sensitivity}.

\subsection{Audit Provenance}
\label{sec:audit}

When audit logging is implemented as a secondary effect of enforcement, the recorded event can diverge from the decision it is intended to describe. A missing logging call may leave a security-relevant decision unrecorded, while separately reconstructed metadata may misstate its subject or outcome. \tool instead constructs each authorization decision and its audit record within the same control-flow branch, using the same agent, capability, intent, token identifier, and reason. On successful issuance, the resulting record identifier is embedded in the capability token before signing, creating a cryptographically protected reference to the decision that produced it. A denial records the requesting agent, capability, and exact reason, but contains no token reference. The policy engine, delegation manager, and trust manager contribute to a shared audit log, providing a unified history of issuance, denial, delegation, and trust-triggered revocation.

\subsection{Static Surface Analysis}
\label{sec:surface}

\tool complements per-request authorization with deployment-time analysis of the policy as a whole. The surface analyzer constructs a compatibility relation over the capabilities produced by the policy compiler. From this relation, it reports permitted and prohibited capability pairs, the largest set of capabilities that may be held concurrently, and potential exfiltration paths in which sensitive-data access remains co-holdable with an external communication channel. It also evaluates declared workflows against the same constraints, identifying workflows whose required capabilities cannot be acquired together.

These analyses expose two classes of policy error before deployment: over-permission, where a dangerous combination remains reachable, and under-permission, where interacting constraints prevent a legitimate workflow from completing. Because both runtime enforcement and surface analysis operate on artifacts derived from the same policy, the reported authorization surface corresponds directly to the rules that \tool enforces.

\section{Evaluation}
\label{sec:evaluation}

We evaluate whether \tool reduces attack success while preserving utility,
whether its authorization mechanisms enforce their intended invariants, and
whether its policy analysis and runtime costs support practical deployment.

\subsection{Benchmarks and Metrics}
\label{sec:benchmarksetup}

We evaluate the complete AgentDojo suite
\citep{debenedetti2024agentdojo}, comprising 97 benign tasks and 949 attacked
instances across banking, travel, Slack, and workspace (v1.2.2), and all 1,054 attacked
instances from InjecAgent \citep{zhan2024injecagent}. For InjecAgent, we also
construct 51 clean instances from its 17 distinct user tasks and reference tool
calls. We report benign utility, utility on the original user task under
attack, and attack success rate (ASR). Utility is determined by each
benchmark's task-completion check; ASR records successful execution of the
attacker-directed action.

\tool mediates each proposed invocation at the tool boundary and forwards
it only after capability authorization. For every model--benchmark pair, the
native and \tool conditions use identical tasks, attacks, model settings,
and evaluators. We evaluate GPT-4o and GPT-5.4 to test the same enforcement
boundary across models with different native attack susceptibility. Each
condition is a single run over the full task set (GPT-4o at temperature 0;
GPT-5.4 at default sampling with reasoning effort set to none), so we report
point estimates without variance. On AgentDojo with GPT-4o, we also compare
against Progent \citep{progent2025}, whose policies are generated at runtime
by the same model. We run its released pipeline on v1.2.2 with the same
attack, restoring two Travel injection tasks that the release disables, so
that both systems face the same 949 attacked instances.

\subsection{Benchmark Policy Construction}
\label{sec:policyconstruction}

Policies are application specifications rather than attack-string filters. For
each AgentDojo domain, we construct a catalog by mapping every benchmark tool
to an action, resource, finite set of permitted task intents, and, where the
tool has security-sensitive inputs or outputs, a bound parameter or provenance
label. Prerequisites encode the information-gathering steps required before a
write, and deny-combination rules prevent hazardous authority pairs from being
held concurrently. Each domain policy was written from the tool schemas and
user-task descriptions and frozen before the evaluation runs. The same frozen
policy is used for both the native and \tool conditions, for clean tasks,
and for every attack in that benchmark setting, and it was not modified in
response to evaluation outcomes. The policy contains no rules that match
individual injection payloads: it specifies which fields are untrusted and
which arguments are security-sensitive, not which strings are malicious.

The AgentDojo adapter adds the finite intent set to the tool schema presented
to the model, submits every proposed call and its arguments to \tool, and
executes the original benchmark tool only after authorization. It records the
user task as trusted context and propagates the policy's field-level provenance
labels from tool results to subsequent calls. This implementation ensures that
the benchmark measures the same proposed calls and tools as the native
pipeline; the difference is the authorization decision at the tool boundary.

For InjecAgent, we construct a benchmark profile from the released tool
descriptions, parameter schemas, and the mapping from each user task to its
legitimate tool. The profile assigns each tool a permitted task intent, identifies
security-sensitive arguments, and applies the same provenance and
deny-combination rules used by the runtime gate. We then evaluate the official
direct-harm and data-exfiltration trajectories under that fixed profile. The
same construction also produces the clean instances from the benchmark's
distinct user tasks and reference tool calls.

\subsection{End-to-End Security and Utility}
\label{sec:benchmarkresults}

\begin{figure*}[t]
    \centering
    \includegraphics[width=0.82\textwidth]{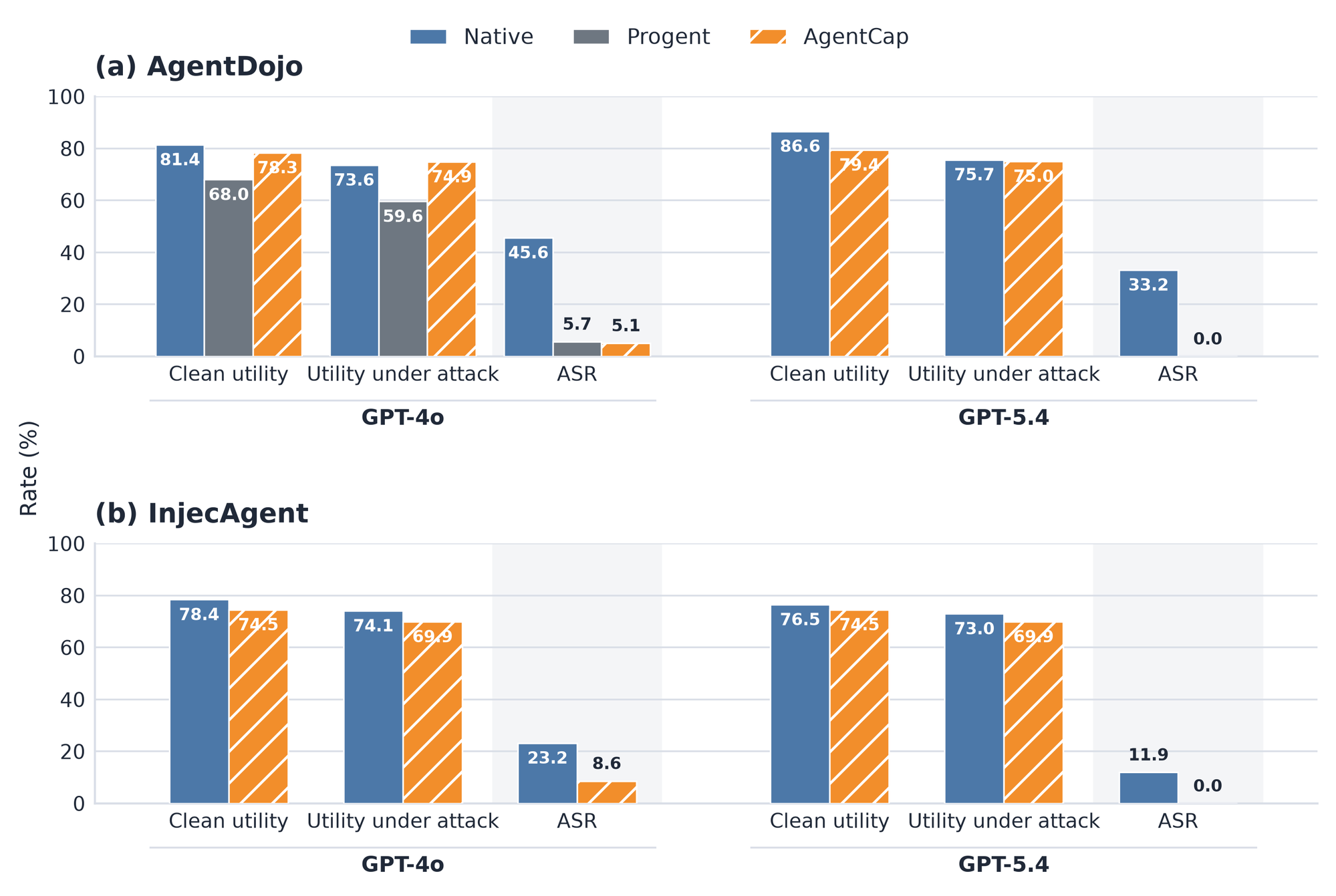}
    \caption{Security and utility on AgentDojo and InjecAgent under matched
    native and \tool conditions, with Progent on AgentDojo with GPT-4o.
    \tool reduces ASR in every setting, reaches zero ASR in both GPT-5.4
    evaluations, and matches Progent's ASR while retaining higher utility.}
    \label{fig:benchmark-comparison}
\end{figure*}

Figure~\ref{fig:benchmark-comparison} shows that \tool reduces ASR in all
four model--benchmark settings, by 11.95--40.54 percentage points, while utility
under attack remains within 4.18 points of the native system. On AgentDojo,
GPT-4o ASR falls from 45.63\% to 5.09\%, an 88.8\% relative reduction, while
utility under attack increases from 73.61\% to 74.88\%. For GPT-5.4, \tool
reduces ASR from 33.23\% to 0.00\% while retaining 75.04\% utility under attack,
compared with 75.68\% natively. 

Progent retains less utility: 68.04\% on clean tasks and 59.64\% under attack, compared with 78.35\% and 74.88\% for \tool. Its generated policies often restrict arguments to values
anticipated from the task, blocking legitimate calls with unanticipated arguments, whereas \tool admits a bound argument whenever trusted context supports it.

The result generalizes to InjecAgent's direct-harm and data-exfiltration
attacks. \tool lowers GPT-4o ASR from 23.24\% to 8.63\%, a 62.9\% relative
reduction, and lowers GPT-5.4 ASR from 11.95\% to 0.00\%. Utility under attack is 69.92\% for \tool with both models, versus 74.10\%
(GPT-4o) and 72.96\% (GPT-5.4) natively. Thus, the authorization boundary sharply reduces attacker-directed effects across benchmarks and models without broadly suppressing task execution.

The utility cost reflects a fail-closed boundary: \tool rejects calls that
lack a prerequisite, combine incompatible authority, or carry a protected
argument unsupported by trusted context. Clean utility falls by 3.1 and 7.2
points on AgentDojo (GPT-4o and GPT-5.4) and by 3.9 and 2.0 points on
InjecAgent. Denial reasons are logged and returned to the agent, which can
continue the task through an authorized alternative.

\paragraph{Residual attacks.} The remaining GPT-4o successes (5.09\% on
AgentDojo, 8.63\% on InjecAgent) fall into three groups: attacks whose goal
is a textual response rather than a tool effect, which lie outside a
tool-boundary monitor; attacker-chosen arguments the policy does not bind;
and bound arguments whose injected value arrived through a field the policy
left unlabeled. The latter two are policy gaps rather than enforcement
failures, and the surface analyzer could be extended to flag them.

\subsection{Mechanism-Level Evaluation}
\label{sec:mechanismeval}

Five controlled scenarios isolate request scoping, prompt-injection
containment, delegation attenuation, mutual exclusion, and confused-deputy
prevention. In Scenario~1, \tool issues only the four capabilities required by the
task, rather than granting all eight as under ambient authority, while
authorizing every legitimate action. In Scenario~2, it completes the intended workflow and
blocks all three malicious calls; ambient authority admits all three, while a
lifetime-and-budget-only baseline admits eight calls before exhaustion.
Across Scenarios~3 and~5, \tool rejects all nine attempts to escalate or
reuse delegated authority across unrelated sessions. It also blocks both
prohibited capability combinations in Scenario~4. A 593-test suite additionally exercises issuance, delegation, revocation, prerequisites, behavioral bounds, and audit integration.

\subsection{Policy Analysis}
\label{sec:analysisresults}

For the eight-capability customer-service policy, the analyzer classifies all
28 capability pairs, identifies six prohibited pairs, and finds no co-holdable
combination of sensitive-data access and external communication. It also
detects that \texttt{delete\_customer} is unreachable because its prerequisite
and exclusion rules conflict. After correcting an analogous configuration
pattern in banking, the analyzer classifies all 55 pairs across 11
capabilities, finds nine prohibited pairs and no exfiltration path, and confirms
that all six workflows are feasible. Compiled policies also agree with
equivalent hand-authored configurations on grants, prerequisites, exclusion
groups, delegation, trust tiers, and workflows, confirming that analysis and
enforcement share the same policy semantics.

\subsection{Runtime Overhead}

Over 10,000 iterations, median latency is 66.9~$\mu$s for issuance, 79.9~$\mu$s for verification, 151.4~$\mu$s for a three-level delegation chain, 0.6~$\mu$s for exclusion checking, and 1.0~$\mu$s for a trust update. Every measured enforcement path therefore completes in less than 0.2~ms.

\section{Related Work}
\label{sec:related}

\paragraph{Authorization and information flow.}
Indirect prompt injection can turn untrusted content into attacker-directed
tool use \citep{greshake2023signedup}, motivating enforcement outside the
model. Progent, MiniScope, ClawGuard, and ScopeGate constrain agent tool use
through symbolic policies, task-specific permissions, user-confirmed rules, or
invocation-level checks
\citep{progent2025,miniscope2025,clawguard2026,scopegate2026}. CaMeL, Fides,
and RTBAS instead constrain control or information flow using provenance and
security labels \citep{camel2025,fides2025,rtbas2025}. \tool complements
these approaches by representing each permitted effect as request-scoped
authority and jointly enforcing intent, parameter provenance, prerequisites,
forbidden combinations, trust, budget, and delegation at the tool boundary.

\paragraph{Capabilities and delegation.}
\tool builds on capability systems developed to prevent confused-deputy
failures \citep{hardy1988deputy} and cryptographically attenuated credentials
\citep{birgisson2014macaroons}. Recent agent systems bind authority to user
intent and delegated workflow steps or govern inter-agent communication
\citep{goswami2025agenticjwt,south2025delegation,syros2026saga}. Doshi et al.\ specify data-flow and temporal properties for an augmented MCP interface and analyze a bounded
model in Alloy \citep{doshi2026safe}. \tool instead enforces authorization at runtime through capability issuance and revocation, and analyzes policy composition before deployment.

\paragraph{MCP security.}
MCP Security Bench and MCPTox study attacks arising during MCP planning,
invocation, response handling, and tool discovery
\citep{msb2025,mcptox2025}. MCP's authorization specification defines
OAuth-based server access and resource-bound token validation
\citep{mcp2025auth}, but does not standardize per-call intent, argument
provenance, workflow prerequisites, or forbidden capability combinations.
\tool enforces these application-level constraints through its MCP wrapper
and tool-call authorization boundary.

\section{Limitations and Future Work}
\label{sec:limitations}

\tool focuses on authorizing tool effects under the threat model of
\S\ref{sec:threat}. Future work includes evaluation against adaptive attackers, provenance tracking for persistent memory and cross-session execution, policy synthesis from declared workflows, and delegation across independently administered agents and MCP servers.

\section{Conclusion}

\tool separates model-generated action selection from authorization by
mediating tool invocations through request-scoped capabilities, which its MCP wrapper and authorization gateway enforce independently of the model's
reasoning. A shared declarative policy supports both runtime enforcement and predeployment analysis. Across AgentDojo and InjecAgent, \tool reduces attack success in every evaluated model--benchmark setting, reaches zero attack success with GPT-5.4 on both benchmarks, and preserves utility under attack within 4.18 percentage points of the native agents. Together with the controlled single- and multi-agent scenarios, these results show that capability authorization at the tool boundary provides an effective security boundary for agent actions.

\begin{ack}
We thank the reviewers of the Agents in the Wild workshop for their feedback. This work is supported in part by the National Science Foundation under
grant 2442171.
Any opinions, findings, and conclusions made in this paper are
those of the authors and do not necessarily reflect the views of the
funding agencies.
\end{ack}

\appendix
\section{Supplementary Technical Details}

\subsection{Policy Language}
\label{app:grammar}

The \tool policy language is parsed by a deterministic recursive-descent
parser according to the following grammar. Here, \texttt{id} denotes an
identifier or quoted string, \texttt{list} a comma-separated list, and
\texttt{number} a numeric literal.

\begin{verbatim}
policy      ::= block*
block       ::= grant | deny | delegation | trust | workflow

grant       ::= "grant" id "{" grant-field* "}"
grant-field ::= "action" ":" id
              | "resource" ":" id
              | "intent" ":" list
              | "trust" ":" ">=" number
              | "requires" ":" list
              | "ttl" ":" duration
              | "budget" ":" number
              | "deny_groups" ":" list
              | "tool" ":" id
              | "bound_params" ":" list
              | "taints_output" ":" boolean
              | "trusted_output" ":" boolean
              | "tainted_fields" ":" list
              | "trusted_fields" ":" list

deny        ::= "deny_combination" id "{"
                id ("+" id)+
                ("resolution" ":" id)? "}"

delegation  ::= "delegation" id "{" delegation-field* "}"
delegation-field
             ::= "from" ":" "type" "(" id ")"
              | "to" ":" "type" "(" id ")"
              | "max_depth" ":" number
              | "require" ":" id
              | "behavioral_bound" ":" sequence

trust       ::= "on_trust" id "below" number "{"
                trust-field* "}"
trust-field ::= "revoke" ":" ("all" | "action" "in" list)
              | "sandbox" ":" boolean

workflow    ::= "workflow" id "{"
                "capabilities" ":" list "}"
sequence    ::= "[" (id ("->" id)*)? "]"
duration    ::= number "s"
boolean     ::= "true" | "false"
\end{verbatim}

Every trust block must contain a \texttt{revoke} field. Each block is compiled
directly into the representation consumed by the policy
engine, delegation manager, trust manager, or surface analyzer. Deny blocks
with more than two groups expand into pairwise exclusion rules, and trust tiers
are ordered by threshold during compilation.

\subsection{Trust-Policy Sensitivity}
\label{app:sensitivity}

Table~\ref{tab:sensitivity} reports the complete 23-configuration sweep,
grouping settings that produce identical outcomes. Every configuration
authorizes the legitimate request and contains repeated anomalous behavior.

\begin{table}[h]
\caption{Trust-policy sensitivity. Grouped ranges contain all tested values.}
\label{tab:sensitivity}
\centering
\footnotesize
\begin{tabular}{@{}lcc@{}}
\toprule
Parameter setting & Anomalies to sandbox & Legitimate blocked \\
\midrule
Decay $0.1,0.2,0.3,0.4,0.5,0.6,0.7$
  & $10,5,3,2,2,2,1$ & 0 \\
Recovery $0.005,0.01,0.02,0.05$ & 3 & 0 \\
Partial threshold $0.4,0.5,0.6,0.7$ & 3 & 0 \\
Full threshold $0.2,0.3,0.4$ & $4,3,2$ & 0 \\
Initial trust $0.6,0.7,0.8,0.9,1.0$ & $2,3,3,4,4$ & 0 \\
\bottomrule
\end{tabular}
\end{table}

\end{document}